      [1999/12/01 v1.4c Il Nuovo Cimento]
\documentclass[varenna]{cimento}

\usepackage{graphics}
\usepackage{epsfig}
\usepackage{amsfonts}
\usepackage{amsmath}
\usepackage{amssymb}
\usepackage{bm}
\newcommand{\nOneD}{n_\mathrm{1d}}

\newcommand{\Fig}[1]{Figure~\ref{fig:#1}}
\newcommand{\fig}[1]{figure~\ref{fig:#1}}
\newcommand{\sect}[1]{section~\ref{sec:#1}}

\title{Does an isolated quantum system relax?}
\author{B.~Rauer,
T.~Schweigler,
T.~Langen\thanks{current address: JILA, NIST \& Department of Physics, University of Colorado, Boulder, Colorado 80309, USA, tim.langen@colorado.edu},
J.~Schmiedmayer}
\institute{Vienna Center for Quantum Science and Technology, Atominstitut, TU Wien, Stadionallee 2, 1020 Wien, Austria}

\begin{document}

\maketitle

\begin{abstract}
Statistical mechanics is one of the most comprehensive theories in physics. From a boiling pot of water to the complex dynamics of quantum many-body systems it provides a successful connection between the microscopic dynamics of atoms and molecules and the macroscopic properties of matter. However, statistical mechanics only describes the thermal equilibrium situation of a system, and there is no general framework to describe how equilibrium is reached or under which circumstances it can be reached at all. This problem is particularly challenging in quantum mechanics, where unitarity appears to render the very concept of thermalization counterintuitive. With the rapid experimental progress in the control and probing of ultracold quantum gases this question has become within reach of detailed experimental investigations. In these notes we present a series of experiments with ultracold one-dimensional Bose gases, which provide novel insights into this fundamental question. 
\end{abstract}

\section{Motivation}
Does an isolated quantum many-body system that is prepared in a non-thermal initial state relax to thermal equilibrium?  As we know from our everyday experience, many physical systems can very successfully be described by a thermal state. On the other hand, the time-reversal symmetry that results from the unitarity of quantum mechanics seems to make the relaxation to thermal states impossible in an isolated system~\cite{Polkovnikov11}. This seemingly simple question thus addresses the fundamental relation between the macroscopic description of statistical mechanics and the microscopic quantum world. It has been highly contested since the 1920s~\cite{Neumann29} and important theoretical advances have been achieved over the years~\cite{Srednicki94,Rigol2008,Polkovnikov11,Eisert2014}. Variations of this question play important roles in such diverse fields as cosmology, high-energy physics and condensed matter~\cite{Kofman1994,Podolsky2006,Braun-Munzinger2001,Berges2004,Eckstein2009,Moeckel2010}. However, it has only been through the recent experimental progress in manipulation of ultracold quantum gases that this question has become within reach of detailed experimental investigations~\cite{Langen15b}. In the following we will present a series of such experiments, which we performed using ultracold one-dimensional Bose gases. The versatility of these gases allowed us to realize several textbook-like non-equilibrium phenomena, which provide important insights into the dynamics of quantum many-body systems. 

\section{One-dimensional Bose gases}
\label{sec:1d_bose_gases}
Over the last years, one-dimensional (1D) Bose gases have proven to be a versatile testbed for the study of quantum many-body physics in and out of equilibrium. The great interest in these systems stems from several key properties. From the theorist's perspective 1D Bose gases offer a rich variety of interesting many-body effects, while still being tractable with reasonable effort~\cite{Cazalilla2011,Castin04}. On the experimental side their realization using cold atomic gases offers precise control over many system parameters, as well as highly-effective means to probe their dynamics~\cite{Bloch2008}. In this first chapter, we will briefly outline important aspects of the description of 1D Bose gases. For more detailed accounts we refer the reader to Refs. \cite{Cazalilla2011,LangenThesis,Schaff14}.

The experimental realization of a 1D Bose gas follows the familiar procedure based on laser and evaporative cooling that is also used for the production of Bose-Einstein condensates from three-dimensional (3D) Bose gases \cite{Davis1995b,Anderson1995}. However, creating an effectivly 1D system in a 3D world requires extremely asymetric traps with a very tight confinement in all but one spatial directions. The general aim of this tight confinement is to raise the energy splitting of ground and first excited state in the two tightly-confined directions, such that all relevant energy scales of the trapped gas lie below it. For a harmonic trap this means that the temperature $T$ and the chemical potential $\mu$ fulfill $k_BT,\mu\ll\hbar \omega_\perp$, with $k_B$ denoting Boltzmann's and $\hbar$ the reduced Planck constant. This realizes a situation where the dynamics along the radial directions can be integrated out leaving the dynamics along the weakly confined axial direction described by an effective 1D model. Contact interactions in this 1D model can be parametrized by an effective scattering potential with the interaction strength~\cite{Olshanii1998}
\begin{equation}
g = 2\hbar a_s \omega_\perp.\label{eq:g1D}
\end{equation}
Here, $a_s$ is the s-wave scattering length of the gas. Note that this description assumes that microscopic scattering processes still have a 3D character, which is the case as long as the s-wave scattering length is small compared to the ground state width of the tight radial confinement, i.e. $a_s\ll \sqrt{\hbar/m\omega_\perp}$, with $m$ denoting the mass of the atoms.  Interesting effects like confinement-induced resonances can occur when this assumption is no longer valid~\cite{Olshanii1998,Haller2010}. 

Such highly-anisotropic trap configurations can be created in strongly-focussed optical dipole traps~\cite{Dettmer2001,Billy2008,Serwane2011}, optical lattices \cite{Paredes04,Kinoshita06,Morsch2006,Bloch2008} or in magnetic micro traps \cite{Folman2002,Reichel2011}. In our experiments, we rely on the latter because micro traps, as we will see below, allow for a particularly precise and convenient preparation of non-equilibrium states. Typical trap frequencies in our setup are $\omega_\perp = 2\pi\cdot 2\,$kHz in the tightly-confining radial directions and $\omega_\mathrm{ax} = 2\pi\cdot 10\,$Hz in the weakly-confining axial direction. The 1D Bose gas is created in this trap by evaporative cooling of an elongated 3D thermal cloud through the condensation crossover and then further into the 1D regime.

While the preparation of an ultracold 1D Bose gas is similar to the one of an ultracold 3D Bose gas, significantly different physics arise once the gas enters the 1D regime. The Mermin-Wagner theorem~\cite{Mermin1966} tells us that no off-diagonal long-range order can emerge due to the enhanced role of fluctuations in 1D. Consequently, there is no macroscopic occupation of the lowest momentum mode even at $T=0$. Thus no true Bose-Einstein condensation is possible. Instead a large number of distinct degenerate regimes emerges \cite{Petrov2000,Kheruntsyan2003}, which might or might not share some of the familiar features of a Bose-Einstein condensate. 

In the homogeneous limit the system is described by the Lieb-Lininger Hamiltonian~\cite{Lieb63}
\begin{align}
\hat H = \frac{\hbar^2}{2m} \int dz&\,\frac{\partial\hat\Psi^\dagger(z)}{\partial z}\frac{\partial \hat\Psi(z)}{\partial z}\,+\nonumber\\
&+\frac{g}{2}\int dz\, dz^\prime\, \hat\Psi^\dagger(z)\hat\Psi^{\dagger}(z^\prime)\delta(z-z^\prime)\hat\Psi(z^\prime)\hat\Psi(z)\label{eq:LiebLiniger},
\end{align}
where the $\hat\Psi(z)$ denote bosonic field operators. The Lieb-Lininger Hamiltonian is a prime example of a so called integrable model \cite{Lieb63,Lieb63b,Yang69,SutherlandBook}. Such models are characterized by a large number of conserved quantities and have historically been an important topic in mathematical physics. Experiments with 1D Bose gases can thus provide a link between the corresponding deep mathematical insights and physical reality. Most notably, the conserved quantities have a profound influence on the non-equilibrium dynamics of these systems, which makes them particularly interesting for the study of relaxation and thermalization processes~\cite{Rigol2007,Caux13}. 

The interaction strength in Eq.~\eqref{eq:LiebLiniger} can be parameterized by the Lieb-Lininger parameter $\gamma = m g/\hbar^2\nOneD$. Notably the interaction strength increases for decreasing particle densities $\nOneD$. For $\gamma \gg 1$ the gas is in the strongly-interacting Tonks-Girardeau regime~\cite{Paredes04,Kinoshita04}. All experiments presented in these notes are performed with $\gamma \ll 1$, where the gas is a weakly interacting quasi-condensate. In this regime density fluctuations are suppressed and the density distribution is similar to the Thomas-Fermi profile of a BEC. However, the phase fluctuates strongly along the length of the system. 

The suppression of density fluctuations allows us to employ a generalized version of the well-known Bogolibov expansion even though there is no macroscopically occupied mode~\cite{Mora2003}. To that end, we express the field operators in terms of density and phase operators
\begin{equation}
\hat\Psi(z) = e^{i\hat\theta(z)} \sqrt{n_\mathrm{1d} +
\hat n(z)},\label{eq:field_operator}
\end{equation}
which satisfy the bosonic commutation relation 
\begin{equation}
[\hat n(z),\,\hat\theta(z^\prime)] = i\delta(z-z^\prime).\label{eq:commutator}
\end{equation}
Inserting this definition into the Hamiltonian in Eq.~\eqref{eq:LiebLiniger} leads to a quadratic model describing the low-energy limit of the system. The result is known as the Luttinger liquid Hamiltonian
\begin{equation}
	\hat H = \frac{\hbar c}{2} \int dz \bigg[\frac{K}{\pi} \bigg(\frac{\partial\hat\theta(z)}{\partial z}\bigg)^2 + \frac{\pi}{K} \, \hat n(z)^2\bigg] = \sum_k \hbar \omega_k \hat a^\dagger_k \hat a_k.
	\label{eq:luttinger}
\end{equation}
The parameters in this Hamiltonian are the speed of sound $c = \sqrt{g n_\mathrm{1d}/m}$ and the Luttinger parameter $K =\sqrt{n_\mathrm{1d}(\hbar\pi)^2/4gm}$. The corresponding eigenmodes are non-interacting phonons with momentum $k$, linear dispersion relation $\omega_k = ck$ and energies $\hbar \omega_k$. The creation and annihilation operators $\hat a_k$ and $\hat a^\dagger_k$ define the phonon occupation number $\hat n_k = \hat a^\dagger_k\hat a_k$. They are directly related to the Fourier components of density and phase via \begin{align}
 \hat n_k &\sim\left(\hat a_k(t)+\hat a_{-k}^\dagger(t)\right) & \hat \theta_k &\sim\left(\hat a_k(t)-\hat a_{-k}^\dagger(t)\right).
\end{align}
One therefore also speaks of the phase and density quadrature of a phonon. Finally, we note that, besides cold atoms, the Luttinger liquid Hamiltonian also plays an important role in both bosonic and fermionic condensed matter systems~\cite{Bockrath1998,Blumenstein2011,Jompol2009a,Deshpande2010}. 

\section{Creating a non-equilibrium state}
\label{sec:creating_non_equ_state}

As we have already noted above, the main tool in all experiments that are presented in these notes is a magnetic micro trap. This micro trap is realized using an atom chip \cite{Reichel2011}, a collection of current-carrying gold wires, which are micro-fabricated onto a silicon substrate. Apart from the possibility to create traps with the necessary aspect ratio to reach the 1D regime (magnetic field gradients scale as $1/r^2$ with the distance $r$ to the current carrying structure and micro traps allow the positioning of the atoms at very small distances $r\sim100\,\mu$m), the atom chip also allows for a precise dynamical control over the trap parameters. For example, the initial harmonic trap can transversely be transformed into a double well potential. This is realized by radio-frequency (RF) dressing of the magnetic sub-states of the atoms \cite{Schumm2005}. The RF fields are applied through additional wires on the chip, which due to their proximity to the atoms allows for very high RF field amplitudes and a precise control over the field polarization. 

\begin{figure}[tb]
	\centering
		\includegraphics[width=0.85\textwidth]{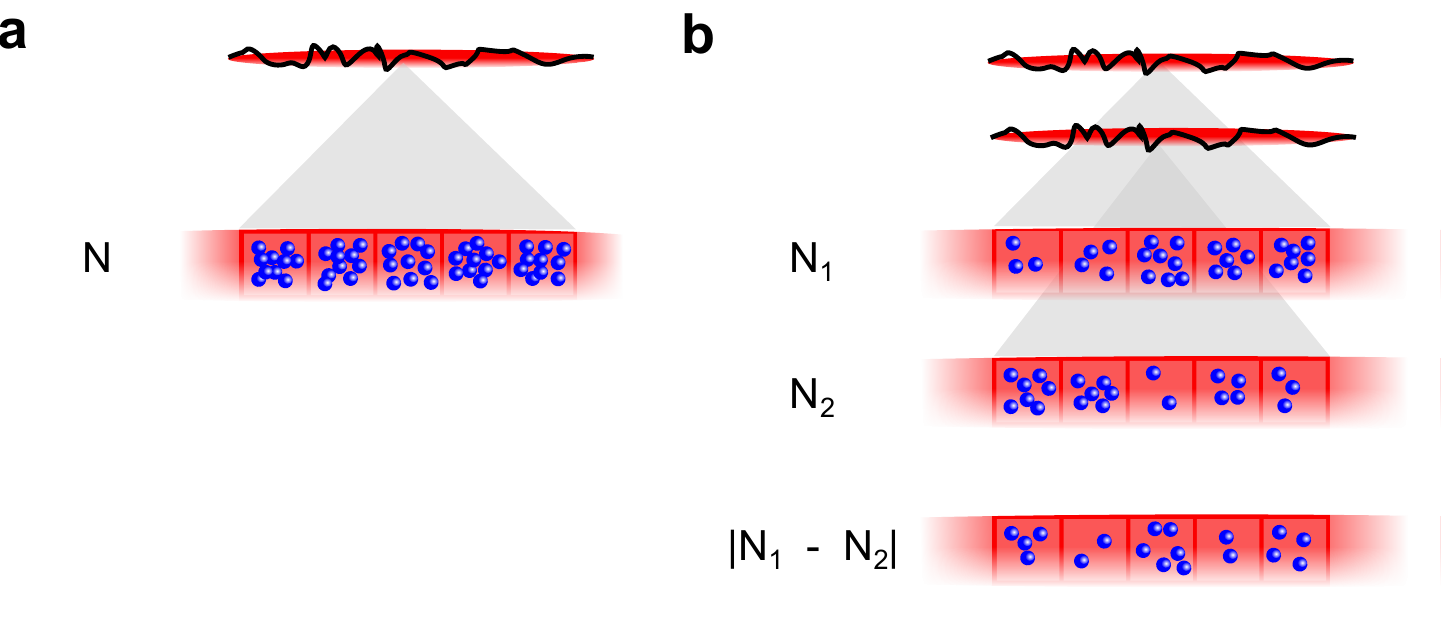}
	\caption{\textbf{Local number fluctuations.} \textbf{(a)} Schematic representation of number and phase fluctuations in a 1D Bose gas. The boxes indicate a course graining on the length scale of the healing length. \textbf{(b)} The splitting distributes the atoms on each of these grid points binomially between the two wells. This results in number fluctuations with a variance of $N/4$ (see text) in each gas. These fluctuations correspond to an energy which is added to the relative degrees of freedom of the system during the splitting. Figure adapted from~\cite{LangenThesis}.}\label{fig:splitting_initial_conditions_binomial_atomnumber}
\end{figure}

We use this technique to coherently split a single 1D Bose gas into two halves, thereby creating a non-equilibrium state~\cite{Kitagawa10,Kitagawa11}. The process of splitting is performed fast compared to the axial dynamics in the system so that $t_\mathrm{split} < \xi_\mathrm{h}/c = \hbar/\mu$. Here $\xi_\mathrm{h} = \hbar/mc$ is the healing length, $c = \sqrt{\mu/m}$ the speed of sound and $\mu$ the chemical potential. The fast splitting assures that no correlations can build up along the axial direction such that the splitting happens independently at each point in the gas. The process can be intuitively pictured as a local beam splitter where each atom is independently distributed into the left or right half of the new system. The corresponding probability distribution for the local number of particles $N$ on each side is therefore binomial  
\begin{equation}
P(N_l,N_r) = \binom{N_l + N_r}{N_l} p_1^{N_l}(1-p_1)^{N_r},
\end{equation}
with $p_1 = 1/2$ for a balanced splitting process. The resulting fluctuations in one half of the system are thus given by $\mathrm{Var} [N_{l,r}] = N \, p_1 \, (1-p_1)$, which translates into $\langle |\Delta N|^2\rangle = N/4$ for $\Delta N = (N_l - N_r)/2$ in the balanced case. \Fig{splitting_initial_conditions_binomial_atomnumber} illustrates this process. 

Once we can speak of two spatially separated systems we can perform a variable transformation to anti-symmetric and symmetric degrees of freedom, which will help us to better describe the quantum state after the splitting. In the following these will also be referred to as relative and common degrees of freedom. Starting from the density and phase fluctuations in the left and right halves (denoted by $\hat n_{l,r}(z)$ and $\hat\theta_{l,r}(z)$, respectively) we find  
\begin{equation}
\hat\phi(z) = \hat\theta_r(z)-\hat\theta_l(z) \quad\textrm{,}\quad \hat\phi_\mathrm{com}(z)=\frac{\hat\theta_r(z)+\hat\theta_l(z)}{2}
\label{eq:relativephase}
\end{equation}
for the phase, and 
\begin{equation}
\hat\nu(z)=\frac{\hat n_r(z)-\hat n_l(z)}{2} \quad\textrm{,}\quad \hat\nu_\mathrm{com}(z)=\hat n_r(z)+\hat n_l(z)
\end{equation}
for the density. The usefulness of this approach becomes clear as we return to the shot noise, which now only enters in the relative number fluctuations 
\begin{equation}
\langle\hat\nu(z)\hat\nu(z^\prime)\rangle= \frac{\nOneD}{2} \delta(z-z^\prime).
\end{equation}
Here, $\nOneD$ denotes the mean density in a single gas after splitting, which results in the additional factor of $2$ as compared to the binomial fluctuations that were introduced above. Transforming these fluctuations into momentum space gives
\begin{equation}
\langle{\hat\nu_k\hat\nu_{k^\prime}}\rangle= \frac{\nOneD}{2} \delta_{k,-k^\prime}.\label{eq:densityfluct}
\end{equation}
From the commutation relation in Eq.~\eqref{eq:commutator}, we see that the corresponding shotnoise introduced to the phase quadrature of the relative modes goes with $1/\nOneD$ and is therefore negligible. 

Returning to the Luttinger Hamiltonian (Eq. \eqref{eq:luttinger}) we can identify the amount of energy that is introduced into each individual phononic mode during the splitting process as $g \nOneD/2$, which is typically significantly smaller than the thermal energy of the initial gas.  Moreover, as we have just shown this energy is only stored in the density quadrature of the relative degrees of freedom, while it should be equipartitioned between phase and density quadrature in thermal equilibrium. 

The situation is different for the common degrees of freedom, which inherit all thermal excitations that were present in the initial gas before the splitting. The state created by splitting is thus also out of equilibrium in this respect, as the common degrees of freedom contain a lot of thermal energy, while the relative degrees of freedom only contain quantum shotnoise. 

In experiment, the equilibrium situation can be realized by the transforming the harmonic trap into a double well while the gas is still thermal. Further independent evaporative cooling in both wells then results in two degenerate gases with no knowledge of each other, which corresponds exactly to thermal equilibrium. The experiment thus enables the unique possibility to contrast non-equilibrium and thermal states in identical settings.

\section{Probing the quantum state}
\label{sec:probing_the_quantum_state}

Information about the system and its dynamics after the splitting is extracted using standard absorption imaging~\cite{Ketterle99} after releasing the system from the trap. If only a single gas is present it simply expands in time-of-flight (TOF), while a pair of condensates expands, overlaps and forms a matter-wave interference pattern~\cite{Schaff14}. The resulting cloud is subsequently illuminated by a resonant laser beam, casting a shadow that is imaged onto a CCD camera. This method is destructive, therefore many identical realizations are necessary to probe a time evolution. It is important to note that the tight transversal confinement of the 1D gases leads to a very rapid radial expansion, which results in an immediate dilution of the system. Therefore, interaction effects in the expansion are negligible and the absorption images enable comprehensive insights into the properties of the initial trapped system. 

\begin{figure}[tb]
	\centering
		\includegraphics[width=0.95\textwidth]{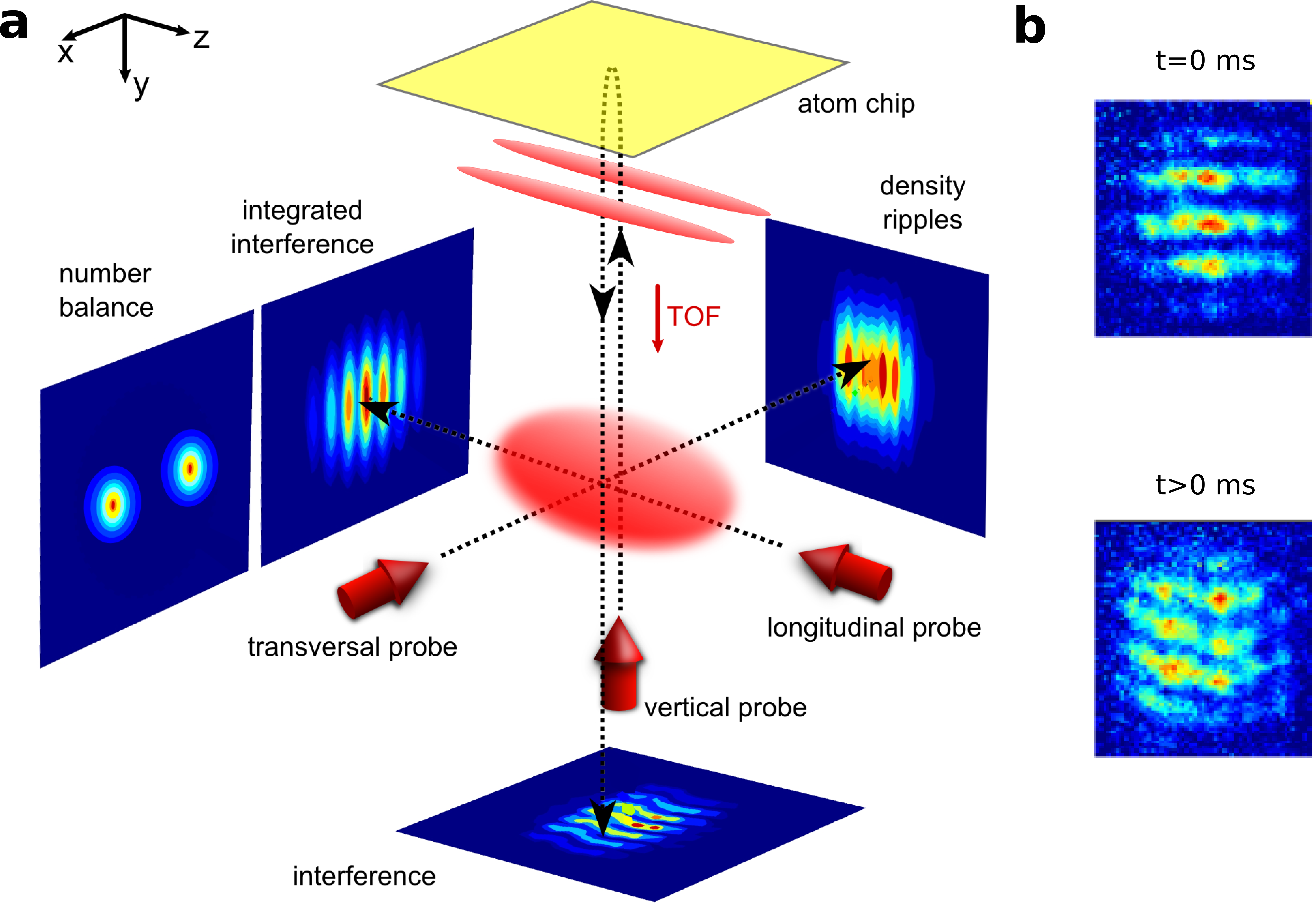}
	\caption{\textbf{(a)} Overview of the available probes in our setup. The transversal probe is primarily used to measure temperature by extracting the density ripple patterns forming in TOF (see \sect{density_ripples}). The vertical probe images the full matter-wave interference pattern containing the entire relative phase field of the two gases (see \sect{full_distribution_functions} and \ref{sec:phase_correlation_functions}). The longitudinal probe records the interference pattern integrated along the 1D direction. It can also be used to measure the number balance by intentionally separating the clouds using a momentum transfer during the trap switch-off. \textbf{(b)} Examples of interference patterns measured with the vertical imaging system right after the splitting ($t=0\,$ms) and after time evolution ($t>0\,$ms). The bending of the fringes reflects the randomization of the relative phase during the dynamics. Figure adapted from~\cite{LangenThesis}. 
}
	\label{fig:imaging_directions}
\end{figure}

A schematic overview of imaging probes employed in our experiment is shown in \fig{imaging_directions}. In the following we will give a short overview of the insights into the dynamics of the quantum state, which are gained through these probes. 

\subsection{Density ripples}
\label{sec:density_ripples}

As we have discussed above, fluctuations play a central role in the physics of 1D Bose gases. It is thus essential that our method allows the probing of a single realization of a 1D Bose gas. In this way, repeating the experiment many times not only gives access to the dynamics but also to the statistical distribution of the fluctuations. It is thus possible to obtain a much deeper insight into the quantum states than would be possible if only mean values of observables could be measured. 

\begin{figure}[tb]
	\centering
		\includegraphics[width=0.85\textwidth]{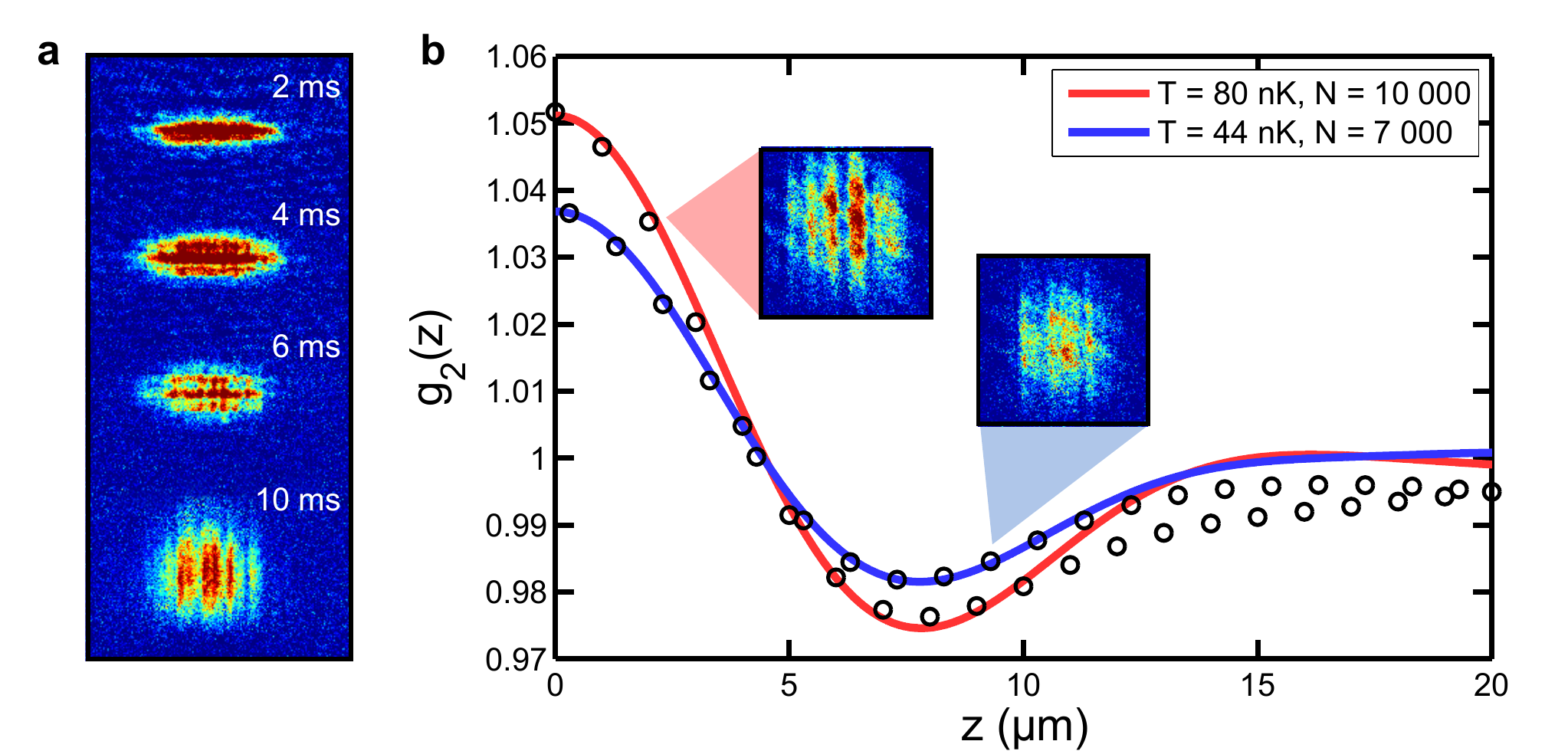}
	\caption{In a TOF measurement the in-situ phase fluctuations transform into density speckle patterns \textbf{(a)}. The correlations in these patterns are used to extract the temperature by fitting them with simulated data \textbf{(b)}. The insets show typical density ripple patterns with the displayed correlations. Figure adapted from~\cite{LangenThesis}.}
	\label{fig:density_ripples}
\end{figure}

A single quasi-condensate that is released and expands in TOF forms strong density speckles along the 1D axis (see \fig{density_ripples}a). These speckles are a direct consequence of the fluctuating phase in the trapped system. In fact, the corresponding gradient $\nabla\theta(z)$ can be interpreted as a velocity field. In expansion this stochastic velocity field transforms into position space realizing a characteristic speckle pattern atop the average density profile. Analyzing the correlations in these patterns and comparing them to simulated results obtained from an Ornstein-Uhlenbeck stochastic process allows us to determine the temperature of the gas~\cite{Imambekov2009,Manz2010} as shown in \fig{density_ripples}b. This is a powerful tool that works as well for 2D systems~\cite{Mazets2012,Langen2013b}. In the experiments it is  primarily used to characterize the initial gas before the splitting. However, it can also be used for the study of the evaporative cooling process~\cite{Grisins2014,Rauer2015} or thermalization (see \sect{long_term_evolution}).

\subsection{Phase correlation functions}
\label{sec:phase_correlation_functions}

The interference pattern of two quasi-condensates as depicted in the lower panel of \fig{imaging_directions} provides a powerful probe for the dynamics of the system. In our case the relative phase fluctuates along the length of the system. In general, the position of the fringes in an interference pattern is determined by this relative phase between the two interfering waves. The meandering fringe pattern in the images thus directly reflects the local in situ relative phase, which can be reconstructed by fitting each local pixel row in the interference pattern with a sinusoidal function.

Right after the splitting the two halves of the system are almost perfectly phase correlated as the shot noise energy is introduced only into the density quadrature, but not the phase. The relative phase is almost zero and the fringes are straight. Over time this coherence is lost and the fringe patterns become more random. This coherence is due to a dephasing of the phononic modes in the relative degrees of freedom. To analyze this process it is instructive to study the correlation function of the relative phase field
\begin{align}
	C(z,z^\prime) &= \frac{\langle\hat\Psi_l{}^\dagger(z)\hat\Psi_r{}(z)\hat\Psi_r{}^\dagger(z^\prime)\hat\Psi_l{}(z^\prime)\rangle}{\langle{|\Psi_r(z)|^2\rangle\langle|\Psi_l(z^\prime)|^2}\rangle} \simeq \langle e^{i\hat\phi(z)-i\hat\phi(z^\prime)}\rangle.
	\label{eq:pcf}
\end{align}
Here, $\hat\Psi_{l,r}$ correspond to the field operators of the left and right gas and $z$ and $z^\prime$ are two points along the axial direction of the system. In the last step we have assumed that density fluctuations can be neglected, which is a very good approximation in the quasi-condensate regime. In the experiment, the expectation value is realized by averaging over many identical realizations. 

For the coherent phase field right after splitting the correlation function is close to one over all relative 
distances $\bar z = z - z^\prime$. After approximately $15\,$ms the systems settles into a steady state, where correlations decay exponentially with $\bar z$. For a 1D Bose gas this exponential decay corresponds to thermal correlations, with the characteristic length scale of the decay $\lambda$ being directly related to the temperature $T$ via $\lambda=\hbar^2 n_\mathrm{1d}/m k_B T$. However, while showing characteristic thermal-like correlations, the relaxed state is markedly different from thermal equilibrium, as its temperature $k_B T_\mathrm{eff} = g n_\mathrm{1d}/2$ can be identified with the shot noise energy that was introduced during the splitting process. Is is thus significantly smaller than the initial temperature $T$ of the system. At the same time, the common degrees of freedom still show a temperature comparable to $T$. The system has thus not fully thermalized, but rather reached a prethermalized state \cite{Gring2012, Kuhnert2013a}, where it already exhibits certain thermal-like features like a temperature. The physical reason behind this is that common and relative degrees of freedom fully decouple in the low-energy limit for a balanced splitting. No energy can be exchanged so that the system can never fully forget its initial state. 

Microscopically this dephasing process can be well understood within the Luttinger description. All energy is initially stored in the density quadrature and all phonons are initialized in phase. During the time evolution the energy of each mode oscillates between density and phase with the momentum-dependent frequency $\omega_k$, which eventually leads to a dephasing. The thermal nature arises from the occupations of the modes. Because of the linear dispersion relation we find that the splitting prepares the relative degrees of freedom with occupation numbers that decay as $1/k$ for increasing momentum $k$. All modes thus obtain the same amount of energy from shot noise, which, after dephasing, makes the state indistinguishable from a thermal state with the corresponding temperature. 

\begin{figure}[tb]
	\centering
		\includegraphics[width=0.60\textwidth]{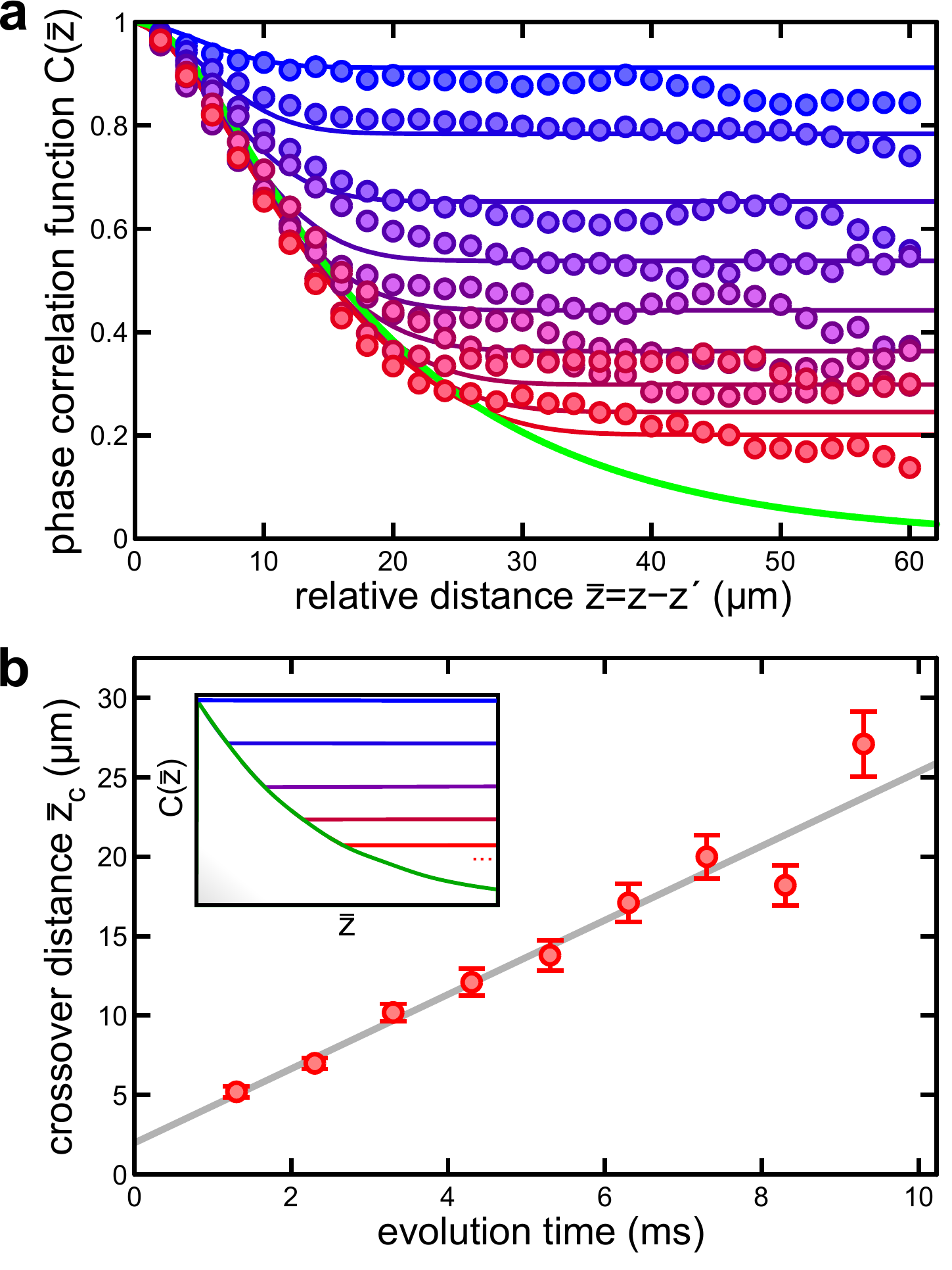}
	\caption{\textbf{(a)} Measured phase correlation functions (circles) of the evolution following the splitting together with the Luttinger liquid predictions (solid lines) taking the trap as well as the finite optical resolution into account. The color encodes time going from blue (1 ms after the splitting) to red (9 ms). The green exponential curve is the prediction for the final prethermalized steady state. \textbf{(b)} Evolution of the crossover distance $z_c$ between the exponentially decaying correlations and the plateau with long range order. The linear behavior shows that the thermal correlations appear locally and spread through the system in a light-cone-like fashion. Figure adapted from~\cite{LangenThesis}.}
	\label{fig:phase_correlation_functions}
\end{figure}

More insights can be obtained by studying the details of the correlation functions during the relaxation process. Their evolution is plotted in \fig{phase_correlation_functions}a~\cite{Langen2013}. For a given point in time the correlations decay exponentially up to a certain crossover distance $z_c$ beyond which the long-range order of the initial state prevails. The evolution of this crossover point plotted in \fig{phase_correlation_functions}b is linear, revealing that the exponentially decaying correlations spread through the system in a light-cone-like dynamic with a characteristic velocity.  This process is driven by the dephasing of the phononic modes of the initial state. Short wavelength modes dephase faster than long wavelength modes leading the characteristic spread. The velocity can be identified with the speed of sound of the phonons, which thus act as carriers of information in the system. This observation provides a direct connection between the establishment of thermal properties and the propagation of correlations in a quantum many-body system. The underlying principles are even more general and also govern the distribution of entanglement, with profound implications, e.g. for quantum information science and computer simulations of complex materials~\cite{Lieb72,Cheneau12,Eisert2010}.   

\subsection{Full distribution functions}
\label{sec:full_distribution_functions}

Another powerful technique to analyze the correlation properties during the relaxation dynamics and especially in the steady state is the full distribution function (FDF) of the interference contrast. To introduce the contrast as an observable we define the operator 
\begin{equation}
	\label{eq:interference_operator}
	\hat A(L)= \int_{L/2}^{L/2} dz\, \hat\Psi_l{}^\dagger(z,t)\hat\Psi_r{}(z,t),
\end{equation}
which corresponds the interference term of the bosonic field operators integrated over a length $L$. The magnitude of $\hat A(L)$ is related to the integrated contrast of the interference patterns $\langle{C^2(L)}\rangle=\langle|\hat A(L)|^2\rangle/\nOneD^2 L^2$. Experimentally the distribution of the squared contrast normalized by the mean squared contrast $\alpha = C^2/\langle|C|^2\rangle$ is less prone to systematic errors and therefore favorable. Recording the shot-to-shot fluctuations of this quantity gives us the full distribution function $P(\alpha)d\alpha$ of the probability to observe a contrast in the interval $\alpha + d\alpha$. The FDFs therefore contain the information about all even moments of the interference operator \eqref{eq:interference_operator} defined above 
\begin{equation}
	\frac{\langle|\hat {A}|^{2m}\rangle}{\langle|\hat {A}|^{2}\rangle^m}=\langle\alpha^m\rangle=\int_0^\infty P(\alpha) \alpha^m d\alpha.
\end{equation}
Thus, they contain much more information about the quantum state than the two-point correlation function introduced earlier.

\begin{figure}[htb]
	\centering
		\includegraphics[width=1\textwidth]{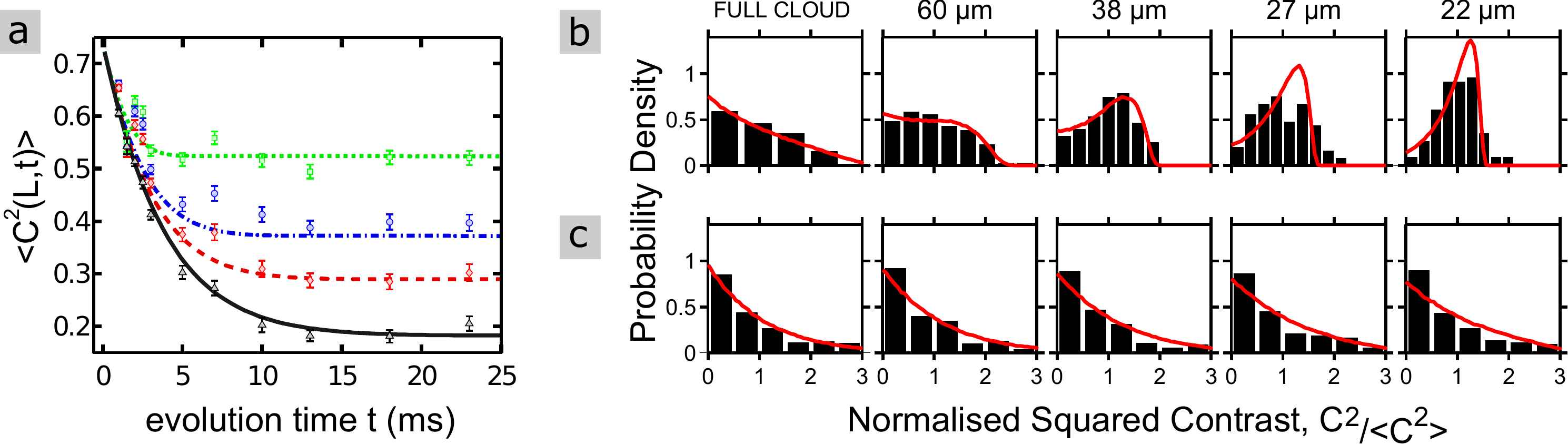}
	\caption{Contrast dynamics and full distribution functions of a coherently split 1D Bose gas. (a) Measured values of the mean squared contrast for various integration lengths $L$ (points). From top to bottom: $L = 18,40,60,100\,\mu$m. The lines show the results of a Luttinger liquid calculation for these integration lengths. (b) Full distribution functions after relaxation to the prethermalized state. The solid red lines show theoretical equilibrium distributions with an effective temperature of $T_\mathrm{eff} = 14\,$ nK, which is significantly lower than the true initial temperature of the gas ($T = 120\,$nK). The prethermalized nature of the state is clearly revealed by comparing it to the vastly different thermal equilibrium situation shown in (c), which can be prepared by creating two completely independent 1D Bose gases. Figure adapted from Refs.~\cite{Gring2012,Kuhnert2013a}.}
	\label{fig:contrast_and_fdfs}
\end{figure}

\Fig{contrast_and_fdfs}a shows the evolution of the mean squared contrast as a function of time, \Fig{contrast_and_fdfs}b a comparison of the FDFs of the prethermalized state discussed in \sect{phase_correlation_functions} and the predictions of the Luttinger liquid model. In \fig{contrast_and_fdfs}c the FDFs of a system of two independent condensates in thermal equilibrium are plotted for comparison. Due to the low effective temperature of the prethermalized state its distributions are peaked over long integration lengths while the much hotter thermal state in \fig{contrast_and_fdfs}c exponentially decays over all observed length scales. This illustrates the fact that the steady state reached after splitting is not the thermal equilibrium of the system.

\section{Generalized Gibbs ensemble}
\label{sec:gge}
The fact that the phonon occupations of the system are preserved during the dynamics is deeply rooted in the integrability of the underlying model. Each relative mode acts like a harmonic oscillator that does not interact but dephases with respect to the rest of the system. This is a general feature of an integrable quantum system, where multiple non-trivial quantities are conserved, severely restricting the system's dynamics. This was strikingly visualized in a landmark experiment by Kinoshita et al.~\cite{Kinoshita06}, which realized the quantum analog of the well known (classical) Newton cradle. Even after thousands of collisions between its constituents such a system will not reach thermal equilibrium, simply because the momenta are conserved and can thus never reach the values given by the Bose-Einstein distribution. 

Nevertheless, it has been conjectured that such systems still relax to a maximum entropy state which is given by the density matrix of a so-called generalized Gibbs ensemble (GGE)~\cite{Rigol2007}
\begin{equation}
	\hat\rho = \frac{1}{Z}e^{-\sum \lambda_j \hat I_j}.
\end{equation}
Here, $Z$ is the partition function, $\hat I_j$ are the operators of the conserved quantities and $\lambda_j$ the corresponding Lagrange multipliers. If only energy is conserved this density matrix reduces to the well-known canonical or Gibbs ensemble, with temperature being the only Lagrange multiplier. If many more conserved quantities exist like the phonon occupations in the Luttinger liquid model, many generalized temperatures, one for each conserved quantity are necessary to maximize entropy. 

\begin{figure}[htb]
	\centering
		\includegraphics[width=1\textwidth]{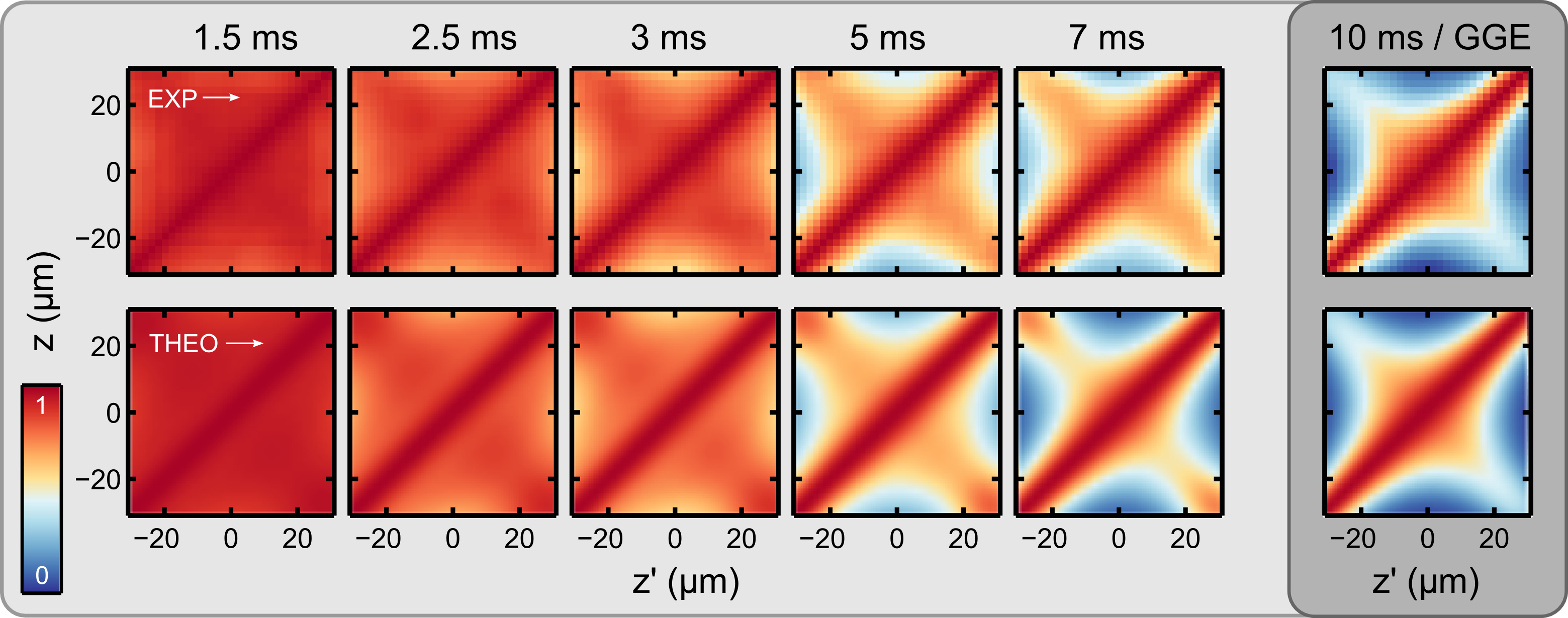}
	\caption{Relaxation dynamics of a coherently split 1D Bose gas with different populations for different modes. Two-point correlation functions $C(z,z')$ for increasing evolution time, showing maxima on the diagonal and the anti-diagonal. The experimental observations (top row) are in very good agreement with the theoretical model (bottom row) demonstrating the presence of many different temperatures in the system. Figure adapted from~\cite{LangenThesis,LangenGGEarxiv}.}
	\label{fig:GGE}
\end{figure}

In our case the occupation numbers of all modes are the conserved quantities. However, the prethermalized state that we have studied so far is a special case of this ensemble, as all temperatures are identical due to the equipartition of energy during the splitting process. To demonstrate the presence of a GGE it is thus necessary to change the splitting process, such that different modes exhibit different temperatures. The results are shown in Fig. \ref{fig:GGE}. Again, the relative phase correlation function can be used to characterize the dynamical states of the system. While we were previously showing only one coordinate of this function, plotting the full function provides straightforward insights into the new occupation numbers. The correlation functions show a trivial maximum on the diagonal ($z_1 = z_2$) which arises due to the fact that every point is perfectly correlated with itself. However, a second maximum arises on the anti-diagonal ($z_1 = -z_2$), indicating that points that are located symmetrically around the center of the system are more strongly correlated. In a simplified model, this implies that modes which are symmetric around the center are more strongly occupied than modes with are anti-symmetric around the center. A more detailed analysis of the relaxed state allows to extract all mode occupations that are necessary to describe the state~\cite{LangenGGEarxiv}. Given these extracted occupation numbers the dephasing model also provides a detailed description of the dynamics, which proves that the conserved quantities were indeed set during the splitting process. 

Most importantly, these observations visualize, both experimentally and theoretically, how the unitary evolution of our quantum many-body system connects to a steady state that can be described by a thermodynamical ensemble.

\section{Dynamics beyond prethermalization}
\label{sec:long_term_evolution}

In sections~\ref{sec:probing_the_quantum_state} and \ref{sec:gge} of these notes we demonstrated that the 1D Bose gases realized in experiment do not relax to thermal equilibrium but to a prethermalized state that can be described by a generalized Gibbs ensemble. This behavior roots in the integrability of the Lieb-Liniger model and its low-energy approximation, the Luttinger liquid model. However, the 1D Bose gas realized in our experiments is only nearly-integrable. On the one hand radial excited states can affect the 1D dynamics and on the other hand the harmonic trap breaks integrability of the Lieb-Liniger model (while integrability is still retained in the trapped Luttinger liquid model~\cite{Geiger2014}). 

It has been conjectured that in this case the observed prethermalized state is only an intermediate steady state on the way to thermal equilibrium, its lifetime being directly related to the degree of integrability breaking~\cite{Kollar11,Stark13}. The analysis of this scenario in the context of classical mechanics has culminated in the important Kolmogorov-Arnold-Moser (KAM) theorem~\cite{Kolmogorov1954}. No complete analogue of this theorem has so far been found in quantum mechanics~\cite{Brandino2014}. Alternatively, also other behaviour has been suggested, namely that the quasi-particles of the experimentally realized 1D Bose gas could be unaffected by the radial states~\cite{MazetsPrivateComm}, leaving the gas fully integrable. Experimental investigations into this effect are ongoing in our and other groups~\cite{WeissPrivateComm}. 

However, even within the coherent dynamics the long-term evolution of the system is expected to show a rich variety of effects, which we will discuss in the following. 

\subsection{Recurrences}
\label{sec:recurrences}

We have shown in the previous chapters that the unitary quantum evolution of a 1D Bose gas can lead to the establishment of thermal properties. This does not mean that a true thermal state was reached, but rather that the expectation values of certain observables became indistinguishable from the corresponding thermal values. In this way the predictions of statistical and quantum mechanics are reconciled. 

However, in a finite system as the trapped system we are dealing with, the unitarity is still expected to result in observable consequences as it forces the dynamics to be periodic. The important question is how long the timescale of this periodic behaviour will be. In the context of our experiments periodic behavior would correspond to a rephasing of the phonons (and thus a reestablishment of coherence) after a finite time, which would be observable as a phase correlation function close to one, $C(\bar z)=1$, over all distances $\bar z$. 

\begin{figure}[htb]
	\centering
		\includegraphics[width=1\textwidth]{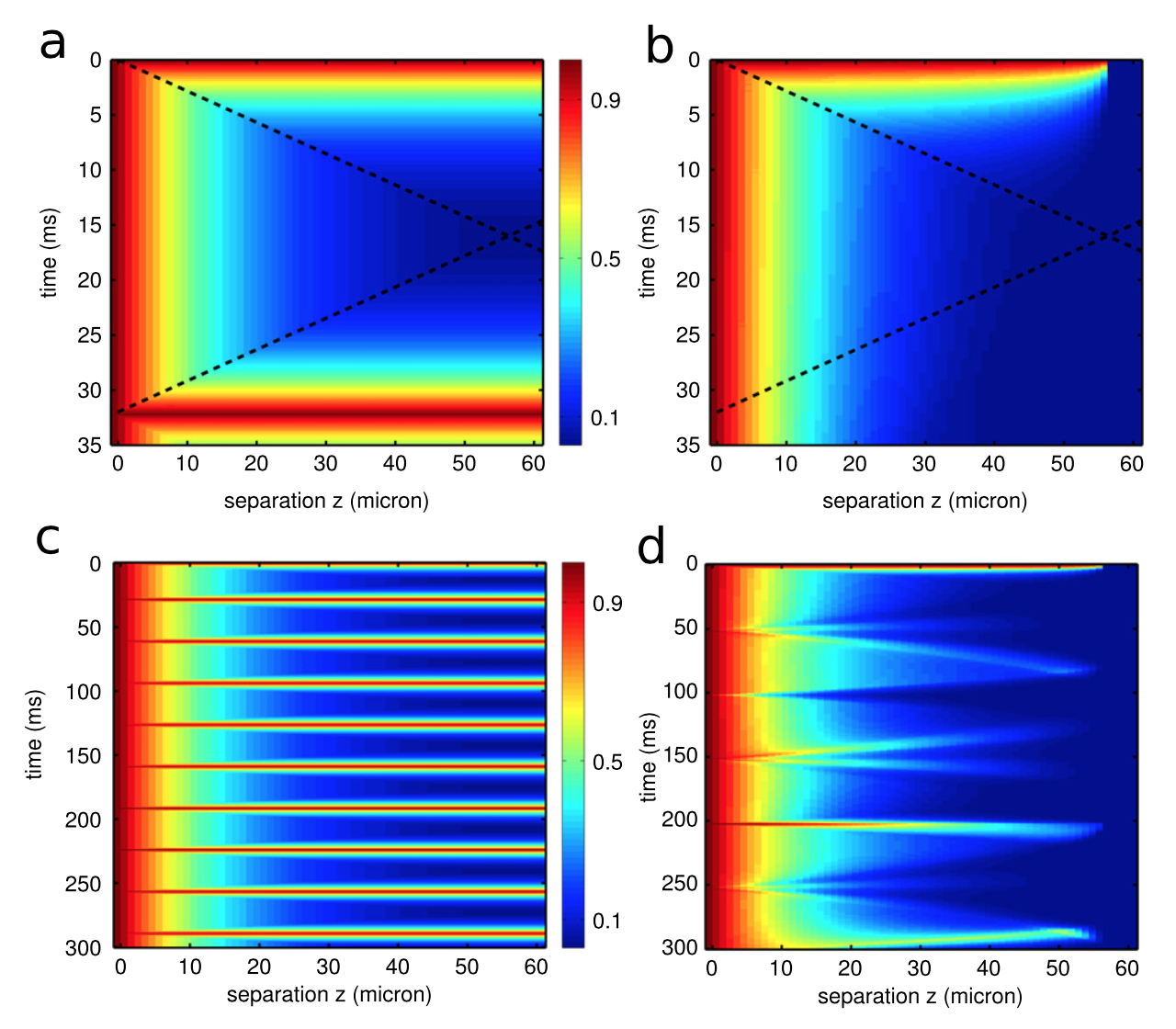}
	\caption{Time evolution of the relative phase correlation function for the homogeneous (left, a and c) and trapped (right, b and d) systems. The color-scale indicates the degree of correlation (red: high correlation, blue: low correlation). The top row illustrates the relaxation to the prethermalized state. In the homogeneous case, the initial state is re-established at times which are multiples of the system length divided by the characteristic velocity. In the trapped case, the recurrences are only partial and the more complex structure is due to the incommensurate ratios of the mode frequencies. In this time window ($0-300\,$ ms), the strongest recurrence is observed at $202\,$ms (Here, $\omega_\mathrm{ax} = 2\pi\cdot 7\,$Hz). Reproduced from ~\cite{Geiger2014}}.
	\label{fig:revival_pcfs}
\end{figure}

In a homogeneous system the time between these recurrences can be estimate as $t_{rec} = L/2c$, which corresponds to twice the time to reach the perfectly dephased prethermalized state. For typical parameters $t_{rec}\sim 30\,$ms. Surprisingly no signs of these recurrences are observed in experiment. 
The reason for this lies in the mode structure of the trapped system. While in the homogeneous case the mode energies are equally spaced $\omega_k = ck$, the modes in a harmonically trapped condensate are described by Legendre polynomials~\cite{Petrov2004}. This leads to the modified dispersion relation
\begin{equation}
	\omega_j = \omega_\mathrm{ax} \sqrt{j(j+1)/2},
\end{equation}
where $\omega_\mathrm{ax}$ is the trap frequency of the axial harmonic confinement and $j$ is the mode index. 
For the given parameters the incommensurate mode frequencies shift the first significant revival in the trapped case to about $200\,$ms, which is challenging to study in experiment~\cite{Geiger2014}. \Fig{revival_pcfs} shows a comparison of the phase correlation dynamics after splitting for the homogeneous and the trapped case. While the initial dephasing dynamics is very similar in both traps the revival structure is quite different, as expected from the dispersion relations. A classical analogy for these dynamics is the behavior of a collection of uncoupled pendula of different length, which only rephase if their frequencies are commensurate. 

To actually measure recurrences in experiment it would be beneficial to trap atoms in a box shaped potential. Flat bottom traps were recently realized for 3D and 2D systems \cite{Gaunt2013, Chomaz2014}. They are a powerful tool to investigate Bose-Einstein condensation or the Kibble-Zurek mechanism unperturbed by trap effects.

\subsection{Imbalanced splitting}
\label{sec:imbalance}

Another relaxation mechanism that is captured by the low-energy description is the dephasing due to imbalances in the splitting process. In practice, the two wells of the double-well potential can never be perfectly balanced during the splitting process. This leads to relative fluctuations of the overall number of atoms in each well. The gas which ends up with more atoms is characterized by a slightly higher chemical potential and speed of sound. These relative differences lead to a dephasing of the two gases with a characteristic velocity $c^\prime = (c_l - c_r)/2$~\cite{LangenThesis,Kitagawa11}. If the atom number difference between the two gases is very small this process will thus be much slower than the initial relaxation to the prethermalized state. However, on long time-scales it will lead to a state in which common and relative degrees of freedom share the same temperature. For an observer, the state will thus be indistinguishable from thermal equilibrium, highlighting again the importance of dephasing and the role of observation for the understanding of thermalization.

\section{Application: Interferometry with squeezed states}

We will end with an application of the well-characterized matter-wave interferometer that we have introduced during the course of these notes. With this we aim to indicate how the fundamental research on non-equilibrium dynamics might also have immediate technological impact in the near future.  

The binomial splitting of a single gas that we discussed above is only a good approximation in the limit of non-interacting atoms. Even for weak interactions, as the ones in the present case for $^{87}$Rb, the splitting has to be very fast to reach the binomial splitting limit. For a slower splitting, interactions will start to play a role and lead to the development of correlations between the atoms. These correlations are a valuable resource for precision measurement devices. 

Experimentally, the speed of the splitting can easily be controlled using the atom chip. While an infinitely fast splitting leads to a relative atom number variance of $N$, the variance resulting from a splitting taking a finite time is reduced by the so-called squeezing factor $\xi_N^2 = \sigma_N^2/N$, where $\sigma_N$ is the standard deviation of the relative number distribution. The slower the splitting the lower the factor $\xi_N$
and thus the stronger the squeezing. The corresponding spin squeezing factor $\xi_S = \xi_N/\langle cos \phi \rangle$ ($\phi$ again denoting the relative phase) can be understood as an entanglement witness, i.e. an observable that signals the presence of genuine multi-particle entanglement~\cite{Sorenson01}. The presence of this entanglement in the states created by the splitting leads to a gain in measurement precision, which can not be achieved with classical states~\cite{Gross10}. 

To actually utilize number squeezing in our setup we need to devise a full interferometric sequence. Apart form the splitting process already described at length we need two further building blocks to achieve this. The first is a mechanism to introduce a relative phase shift between the two arms of the interferometer, which emulates the measurement signal in a possible application. Experimentally we realize this by tilting the double well potential after splitting so that the two gases experience a different gravitational potential and accumulate a phase difference. Varying the time $t_\phi$ the system is kept in this state controls the overall phase shift. As a second building block, we need to employ a recombiner that allows for measurements of the relative atom number in the two arms of the interferometer (in analogy to the second beam splitter in an optical Mach-Zehnder interferometer). This can be achieved by accelerating the two gases onto each other while keeping a barrier between them that is small enough to allow for inter-well tunneling. In this process the relative phase is mapped to a relative population difference, just like in the case of two wave packets that simultaneously impinge on a semi-reflective barrier from different sides. In addition, the relative phase $\phi$ between the two wells can also be measured using the standard matter-wave interference procedure that was already described in \sect{probing_the_quantum_state}.  

\begin{figure}[htb]
	\centering
		\includegraphics[width=1\textwidth]{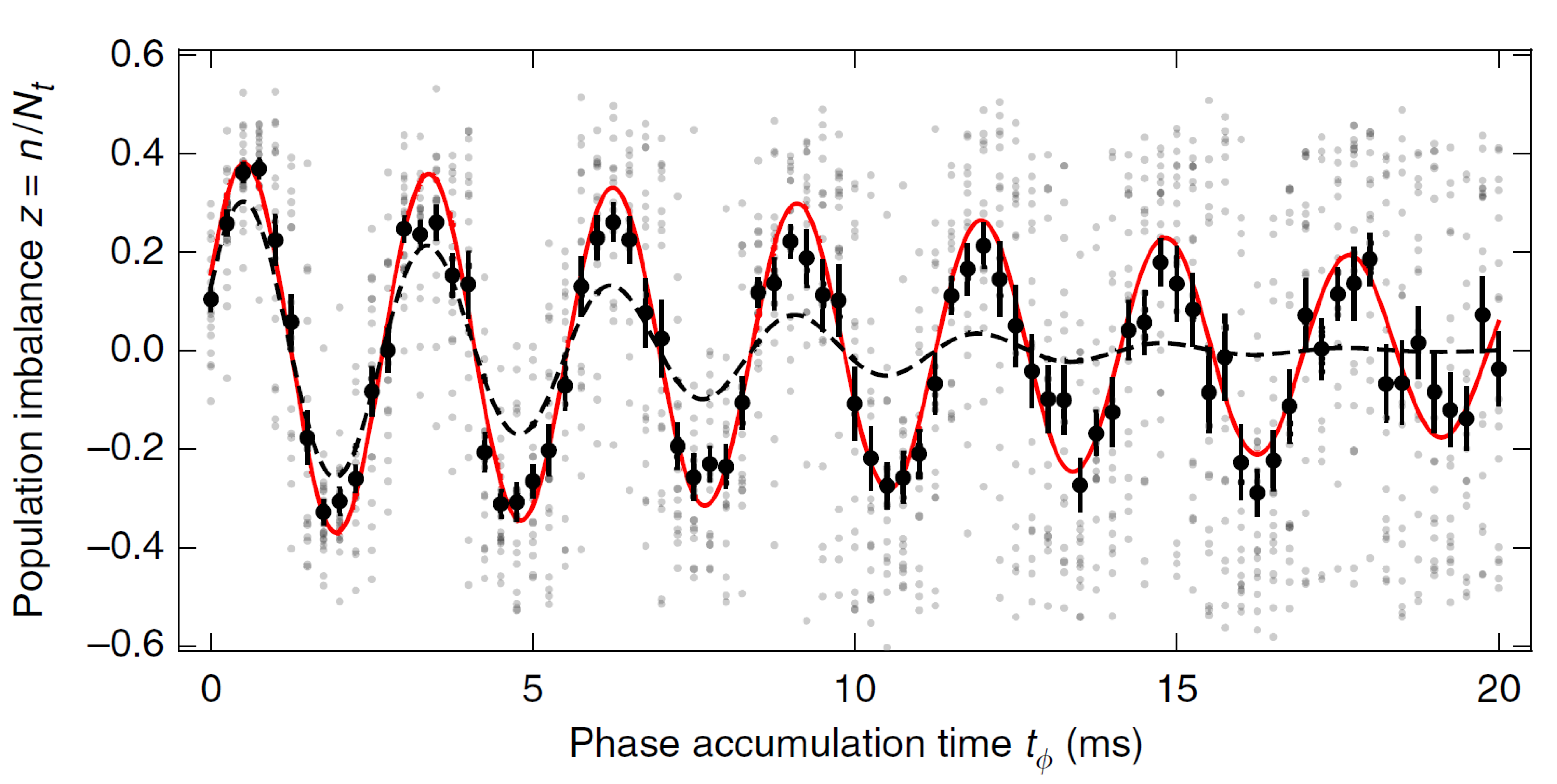}
	\caption{Output signal of the integrated Mach-Zehnder Interferometer. The normalized population difference $z = N_l-N_r/(N_l+N_r)\equiv n/N_t$ between the two wells is measured as a function of time $t_\phi$. It exhibits interference fringes and a damping due to phase diffusion. Grey dots: imbalance of individual experimental realizations; black dots: ensemble average $\langle z \rangle$; red curve: theoretical prediction taking into account phase diffusion; dashed black line: expected signal for a classical coherent state without squeezing. Reproduced with permission from~\cite{Berrada2013}.}
	\label{fig:tarik}
\end{figure}

\Fig{tarik} shows the experimentally observed population imbalance of a squeezed initial state as a function of the phase accumulation time $t_\phi$. The fringe contrast of the average values is damped due to phase diffusion. Naively, this phase diffusion would be expected to be much more severe, but the presence of a long-lived prethermalized state limits its deteriorating effects. Comparing the observed decay time to the one expected for a coherent state of $\xi_N = 1$ (dashed line) illustrates the gain in interferometric precision when using a squeezed input state. The best spin squeezing achieved in this setup is $\xi_S^2 = -7.8 \pm 0.8\,$dB \cite{Berrada2013}, corresponding to genuine multi-particle entanglement of $150$ atoms. This result could in the future be increased by the use of optimized splitting ramps~\cite{Grond2010}, and outlines the way for interferometric sensing of local forces in atom chip configurations. 

\section{Conclusion}
The relaxation of isolated quantum many-body systems is a major unsolved problem connecting statistical and quantum physics. Understanding such relaxation processes remains a challenge despite considerable efforts. 
Experiments with ultracold quantum gases (in general) and 1D Bose gases (in particular) allow the realization and manipulation of well-controlled and truly isolated quantum systems. As we have shown, this provides unique opportunities to study and understand non-equilibrium phenomena. For example, the results discussed in these notes demonstrate for the first time several characteristic aspects of these dynamics, including the existence of a stable, thermal-like prethermalized state and its dynamical, light-cone-like emergence. Furthermore, the connection of the prethermalized state with generalized statistical ensembles, and thus of the unitary quantum evolution and statistical mechanics was highlighted. The progress in this field is rapid and we expect it to continue to have profound implications for our understanding of isolated quantum many-body systems. 

\section{Acknowledgements}
This work was supported by  the  EU  (SIQS  and  ERC  advanced  grant  Quantum-Relax).   B.R.  and  T.S. acknowledge the support by the Austrian Science Fund (FWF) through the Doctoral Program CoQuS (W1210) and through the SFB FoQuS.

\bibliographystyle{bibtex/varenna}
\bibliography{bibtex/Papers-VarennaNotes14_clean,bibtex/add_biblio,bibtex/biblio}
\end{document}